\documentclass[12pt]{article}
\usepackage{graphicx}
\usepackage{amssymb}
\usepackage{amsmath}
\usepackage{cite}

\newcommand{\bea}{\begin{eqnarray}}  \newcommand{\eea}{\end{eqnarray}}
\newcommand{\beq}{\begin{equation}}  \newcommand{\eeq}{\end{equation}}

\newcommand{\zzeta}{{{\cal R}_c}}

\begin{document}

\begin{titlepage}

\begin{flushright}
ICRR-Report-551\\
RESCEU-26/09
\end{flushright}

\vskip 1.35cm

\begin{center}
{\large \bf
Gravitational Wave Background and Non-Gaussianity\\
as a Probe of the Curvaton Scenario
 }
\vskip 1.2cm

Kazunori Nakayama$^{(a)}$ and
Jun'ichi Yokoyama$^{(b,c)}$

\vskip 0.4cm

{ \it $^a$Institute for Cosmic Ray Research (ICRR),
University of Tokyo,
Kashiwa, Chiba 277-8582, Japan}

{ \it $^b$Research Center for the Early Universe (RESCEU), 
Graduate School of Science, The University of Tokyo, Bunkyo-ku,
Tokyo 113-0033, Japan
}

{\it $^c$Institute for the Physics and Mathematics of the Universe (IPMU),
University of Tokyo, Kashiwa 277-8568, Japan}

\date{\today}

\begin{abstract}
We study observational implications of the
stochastic gravitational wave background
and a non-Gaussian feature of scalar perturbations  on the curvaton
mechanism of the generation of density/curvature fluctuations,
and show that they can determine the properties of the curvaton
in a complementary manner to each other.  Therefore even if Planck
could not detect any non-Gaussianity, future space-based laser
interferometers such as DECIGO or BBO could practically exhaust
its parameter space.
\end{abstract}


\end{center}
\end{titlepage}

\section{Introduction}

The idea of inflation \cite{lindebook} has
become a standard paradigm since it was proposed in the early 1980's.
Inflation is an accelerated expansion epoch in the very early 
Universe, which
solves the flatness, the horizon and also the 
monopole problems naturally.
Furthermore quantum fluctuation of the inflaton $\phi$, 
which is a scalar field responsible for the 
accelerated expansion, can provide
the seed of cosmic density/curvature fluctuations \cite{yuragi}
observed through the cosmic microwave background (CMB) 
anisotropy  \cite{Komatsu:2008hk}, galaxy clustering, etc.

Another prediction of inflation, which is in fact more generic,
is the generation of  the stochastic gravitational wave background
or the tensor perturbation,
whose spectrum is nearly scale-invariant \cite{staro,Allen:1987bk,Turner:1990rc,Sahni:1990tx,Turner:1993vb,Turner:1993xz,Liddle:1993zj,Turner:1995ge,Turner:1996ck,Allen:1997ad,Seto:2003kc,
Weinberg:2003ur,Tashiro:2003qp,Ungarelli:2005qb,Smith:2005mm,Seto:2005qy,Kudoh:2005as,Boyle:2005se,Smith:2006xf,Chongchitnan:2006pe,Friedman:2006zt,Zhao:2006is,Watanabe:2006qe,Chiba:2007kz,Boyle:2007zx,jyessay,Smith:2008pf,Nakayama:2008ip,Nakayama:2008wy,
Kuroyanagi:2008ye,Mangilli:2008bw}
(see Ref.~\cite{Maggiore:1999vm} for a review).
The amplitude of the gravitational wave is simply proportional to the
Hubble parameter, $H_{\rm inf}$, during inflation
and many
 inflation models predict detectable amplitude of gravitational waves.
In particular, future space-based gravitational 
wave detectors such as DECIGO \cite{Seto:2001qf}
and/or BBO
have a chance to detect inflationary gravitational wave background.
Recently, it was pointed out that thermal history of 
the early universe is imprinted in
the spectral shape of the gravitational wave background
\cite{Seto:2003kc,Boyle:2005se,Boyle:2007zx,jyessay,
Nakayama:2008ip,Nakayama:2008wy}.
In particular, the reheating temperature of the Universe after inflation
can be determined or constrained from future gravitational wave experiments
\cite{jyessay,Nakayama:2008ip,Nakayama:2008wy}.
Thus any detection of primordial gravitational wave background gives 
useful information
on the early Universe.

Traditionally, the spectrum of the 
comoving curvature perturbation is parametrized as
\begin{equation}
	\Delta^2_{\zzeta}(k)=\Delta^2_{\zzeta}(k_*)\left ( 
\frac{k}{k_*} \right )^{n_s-1}.
\end{equation}
Here $k_\ast$ is the pivot scale which we take
 $k_*=0.002$~Mpc$^{-1}$  and $n_s$ is the scalar spectral index.
Inflation predicts nearly scale-invariant spectrum with $n_s\cong 1$,
with its precise value determined by the shape of the potential \cite{yuragi},
and it agrees well with observations so far~\cite{Komatsu:2008hk}.
If $\phi$ has a single component with a canonical kinetic term,
the resultant curvature fluctuation is Gaussian distributed,
which is again in agreement with the observation today
 \cite{Komatsu:2008hk}.

However, the origin of the density perturbation is not limited to 
the quantum fluctuation of the inflaton.
Another scalar field, called curvaton, may be responsible for the 
generation of the 
observed density perturbation \cite{Mollerach:1989hu,Lyth:2001nq}.
It is undoubtedly an important task to distinguish them in order to 
understand the physics of the
early Universe.
One such signature may come from the non-Gaussianity in the CMB 
anisotropy \cite{Bartolo:2004if},
since the curvaton scenario can produce large enough non-Gaussian 
feature to be detected
\cite{Lyth:2002my,Enqvist:2005pg},
while standard inflation models predict negligible non-Gaussianity
\cite{Acquaviva:2002ud,Maldacena:2002vr,Seery:2005wm,Yokoyama:2007uu}.
However, the situation where the curvaton generates large
non-Gaussianity 
is somewhat limited
and it is possible that the curvaton accounts for the observed 
density perturbation
without generating large non-Gaussianity.

In this paper, we point out that the observation of stochastic 
gravitational wave background plays a very important role to probe the
physics of the curvaton scenario.
In the case that the curvaton generates negligible non-Gaussianity, 
there must be an entropy production process by the curvaton 
decay itself, 
and such a non-standard thermal history is imprinted in the 
spectrum of the 
inflationary gravitational wave background.
Future space-based gravitational wave detectors, such as 
DECIGO 
and/or BBO may be able to do this job.
This opens up a possibility to find an evidence of the 
curvaton scenario.  Thus measurement of the non-linearity parameter 
$f_{\rm NL}$ which is a simplified measure of non-Gaussianity and that of
tensor perturbations by DECIGO/BBO play complementary roles to each
other.

The rest of the paper is organized as follows.
In Sec.~\ref{sec:curvaton}, the gravitational wave background 
spectrum and non-Gaussianity
in the curvaton scenario are summarized.
In Sec.~\ref{sec:probe} the detection possibility is discussed.
Sec.~\ref{sec:conc} is devoted to the conclusion.

\section{Features of the Curvaton Scenario} \label{sec:curvaton}

In this section, we summarize the cosmological 
consequences of the curvaton scenario, namely, the spectra of
gravitational wave background and the density/curvature perturbation,
and non-Gaussianity.
All these ingredients are essential for probing the curvaton 
scenario and 
constraining the parameter space, as will be discussed in 
Sec.~\ref{sec:probe}.

\subsection{Gravitational wave background spectrum in the curvaton scenario}

In the inflationary era, quantum fluctuations of the RMS amplitude of
$H_{\rm inf}/(2\pi)$ 
are induced on all massless fields in each Hubble time.
Here ``massless'' means that the mass is much smaller than $H_{\rm inf}$.
The tensor perturbation of the metric consists of two
 free massless scalar components,
which can be quantized in de-Sitter space-time.
After a mode left the horizon during inflation, 
it actually becomes a classical fluctuation
which can be viewed as a stochastic gravitational wave.
Thus inflation necessarily generates a gravitational wave background
from a cosmological scale to a sub-kilometer scale
The former is a target of detection by the B-mode polarization of
CMB anisotropy and direct detection by space-based laser interferometer
 experiments.

Tensor perturbation of the metric is defined by the following 
line element,
\begin{equation}
	ds^2=a(\tau)^2\left[ -d\tau^2 +(\delta_{ij}+h_{ij})dx^i dx^j \right],
\end{equation}
where $a(\tau)$ is the scale factor and $h_{ij}$ denotes 
the metric perturbation
satisfying the transverse-traceless conditions  $\partial^i h_{ij}=0$
and $h^i_i=0$.
Thus $h_{ij}$ has two physical degrees of freedom, which are denoted 
as $h^\lambda$
with $\lambda=+,\times$.
In the inflationary era, the tensor perturbation has a quantum 
fluctuation
whose spectrum is given by
\begin{equation}
	\Delta_h^{(\rm p) 2}(k)=64\pi G\left 
( \frac{H_{\rm inf}}{2\pi} \right )^2
	\left ( \frac{k}{k_*} \right )^{n_t},
\end{equation}
where the tensor spectral index is given by $n_t=-2\epsilon$. 
Here $\epsilon$ is one of the  slow-roll parameters
during inflation defined by
\begin{equation}
	\epsilon = \frac{M_{\rm Pl}^2}{2}
\left ( \frac{V'}{V} \right )^2,~~~~~
	\eta = M_{\rm Pl}^2 \frac{V''}{V},
\end{equation}
where $V$ is the inflaton potential and the prime denotes 
derivative with respective to
the inflaton field $\phi$, and 
$M_{\rm Pl}=(8\pi G)^{-1/2}$ is the reduced Planck scale.

After the production of the gravitational waves during inflation,
the amplitude of each Fourier mode remains constant when 
the corresponding mode lies outside the Hubble radius.
However, once it enters the horizon, its amplitude 
decreases as $\propto a^{-1}$.
Thus the present energy density of the gravitational wave 
background per logarithmic frequency interval
is written as
\begin{equation}
	\frac{d\rho_{\rm gw}}{d\ln k}=\frac{k^2}{32\pi Ga_0^2}
\Delta_h^{(\rm p) 2}(k)
	\left ( \frac{a_{\rm in}(k)}{a_0} \right )^2,
\end{equation}
where $a_0$ is the present scale factor, and $a_{\rm in}(k)$ 
denotes the scale factor at which
the corresponding mode with wave number $k$ enters the horizon.
It behaves as $\propto k^{-2} (k^0)$ for the mode which enters 
the horizon 
at matter (radiation) dominated era.
Therefore thermal history of the Universe is imprinted in
the spectrum of the gravitational wave background,
and this is the reason why the observations of the gravitational 
wave are expected to
have great impacts on cosmology
\cite{Seto:2003kc,Boyle:2005se,Boyle:2007zx,Nakayama:2008ip,
Nakayama:2008wy}.

In terms of the density parameter, it can be rewritten as
\begin{equation}
	\Omega_{\rm gw}(k)=\frac{k^2}{12a_0^2H_0^2}\Delta_h^2(k),
\end{equation}
where $H_0$ is the present Hubble parameter, and
\begin{equation}
	\Delta_h^2(k) = \Delta_h^{(\rm p)2}(k)
	\Omega_{\rm m} ^{2}
	\left ( \frac{3j_1(z_k)}{z_k} \right )^2
	\left ( \frac{g_*(T_{\rm in})}{g_{*0}} \right )
	\left ( \frac{g_{*s0}}{g_{*s}(T_{\rm in})} \right )^{4/3} 
	T_1^2(x_{\rm eq}) T_2^2(x_{\rm R}),    \label{Deltah}
\end{equation}
where $g_*(T_{\rm in})$ denotes the effective relativistic 
degrees of freedom at the temperature 
$T_{\rm in}$ when $k$-mode enters the horizon,
and $j_1(z)$ is the spherical Bessel function of the 
first rank with $z_k\equiv 2k/(a_0H_0)$.\footnote{
	There was an error in Eq.~(14) of Ref.~\cite{Nakayama:2008wy}.
	Eq.~(\ref{Deltah}) is the correct one.
}
The transfer functions $T_1(x)$ and $T_2(x)$ are given by
\cite{Turner:1993vb,Nakayama:2008wy}
\begin{gather}
	T_1^2(x) = 1+1.57 x +3.42 x^2,\\
	T_2^2(x) = [1-0.32 x +0.99 x^2]^{-1}.
\end{gather}
The former connects the gravitational wave spectrum 
of the mode entering the horizon
before ($x_{\rm eq}\equiv k/k_{\rm eq} >1$) and after 
($x_{\rm eq}<1$) the matter-radiation equality,
where $k_{\rm eq}\equiv a(t_{\rm eq})H(t_{\rm eq})=7.3\times 10^{-2} 
\Omega_{\rm m}h^2~
{\rm Mpc}^{-1}$.\footnote{
	The effect of neutrino free streaming is known to lead
	a suppression on the gravitational wave background 
spectrum around 
	the frequency $\sim 10^{-9}$~Hz 
\cite{Weinberg:2003ur,Watanabe:2006qe}.
	But our concern is around 1~Hz, and 
hence we simply neglect this effect.
}
The latter transfer function 
 connects the mode entering the horizon before 
($x_{\rm R}\equiv k/k_{\rm R} >1$) and after ($x_{\rm R}<1$) 
the reheating subsequent to  inflation.  Here
$k_{\rm R}$ is the comoving wavenumber corresponding to
the horizon scale at the reheating epoch when the 
Universe became radiation dominant. Without any significant
entropy production after the reheating, it is given by
\begin{equation}
\begin{split}
	\frac{k_{\rm R}}{a_0}
&=1.7\times 10^{13}~{\rm Mpc}^{-1}
\left ( \frac{g_{*s}(T_{\rm R})}{106.75} \right )^{1/6}
	\left ( \frac{T_{\rm R}}{10^6~{\rm GeV}} \right ),\\
	&T_{\rm R}=\left(\frac{10}{\pi^2 g_*(T_{\rm R})}\right)^{1/4}
\sqrt{\Gamma_\phi M_{\rm Pl}},  \label{kR}
\end{split}
\end{equation}
with $T_{\rm R}$ and $\Gamma_\phi$ 
being the reheating temperature after inflation and the decay rate of
the inflaton, respectively.  
This corresponds to the frequency
 $f=$0.026Hz for $T_{\rm R}=10^6$GeV, 
which is close to the most sensitive
frequency range of the planned future space-based laser interferometer
experiments, DECIGO or BBO.
Thus by observing the spectral shape of the 
gravitational wave background, 
the reheating temperature of the Universe 
can be determined or constrained.

As mentioned above, the 
above correspondence assumes the standard thermal history with
no significant entropy production after reheating.
In the curvaton scenario, however, its coherent oscillation 
may once dominate the Universe,
which introduces an additional matter-dominated era,
and then decays releasing huge amount of entropy.
In this case the gravitational wave spectrum receives an additional 
suppression 
\cite{Seto:2003kc,Nakayama:2008ip,Nakayama:2008wy}
which can be quantified by the dilution factor, $F$, defined by
\begin{equation}
	F = \frac{s(T_\sigma)a^3(T_\sigma)}{s(T_{\rm R})a^3(T_{\rm R})}
	=\left \{ \begin{array}{ll}
	\displaystyle \frac{\sigma_i^2}{6M_{\rm Pl}^2}\frac{T_{\rm R}}{T_\sigma}
	~~&{\rm for}~~~m_\sigma > \Gamma_\phi \\
	\displaystyle \frac{\sigma_i^2}{6M_{\rm Pl}^2}\frac{T_{\rm osc}}{T_\sigma}
	~~&{\rm for}~~~m_\sigma < \Gamma_\phi
	\end{array} \right.,
	\label{F}
\end{equation}
where $\Gamma_\phi$ is the decay rate of the inflaton,
$T_\sigma$ is the radiation temperature just after the curvaton decay,
and $T_{\rm osc}$ is the temperature at which the curvaton 
begins to oscillate.  This expression is valid for $F \gg 1$.
The resultant gravitational wave spectrum is given by
\begin{equation}
\begin{split}
	\Delta_h^2(k) = &\Delta_h^{(\rm p)2}(k)
	\Omega_{\rm m}^2
	\left ( \frac{3j_1(z_k)}{z_k} \right )^2
	\left ( \frac{g_*(T_{\rm in})}{g_{*0}} \right )
	\left ( \frac{g_{*s0}}{g_{*s}(T_{\rm in})} \right )^{4/3} \\
	&\times T_1^2(x_{\rm eq}) 
	T_2^2(x_{\sigma})
	T_1^2(x_{\sigma \rm R})
	T_2^2(x_{\rm R}(F)),
\end{split}
\end{equation}
where $x_\sigma=k/k_\sigma$ with $k_\sigma$ given in 
an analogous way with (\ref{kR}) after 
replacing $T_{\rm R}$ with $T_\sigma$,
and $x_{\sigma \rm R} = k/k_{\sigma \rm R}$ with 
$k_{\sigma \rm R} = k_\sigma F^{2/3}$
and $x_{\rm R}(F) = k/k_{\rm R}(F)$ with 
$k_{\rm R}(F) = k_{\rm R} F^{-1/3}$.

\begin{figure}[]
 \begin{center}
 \includegraphics[width=0.7\linewidth]{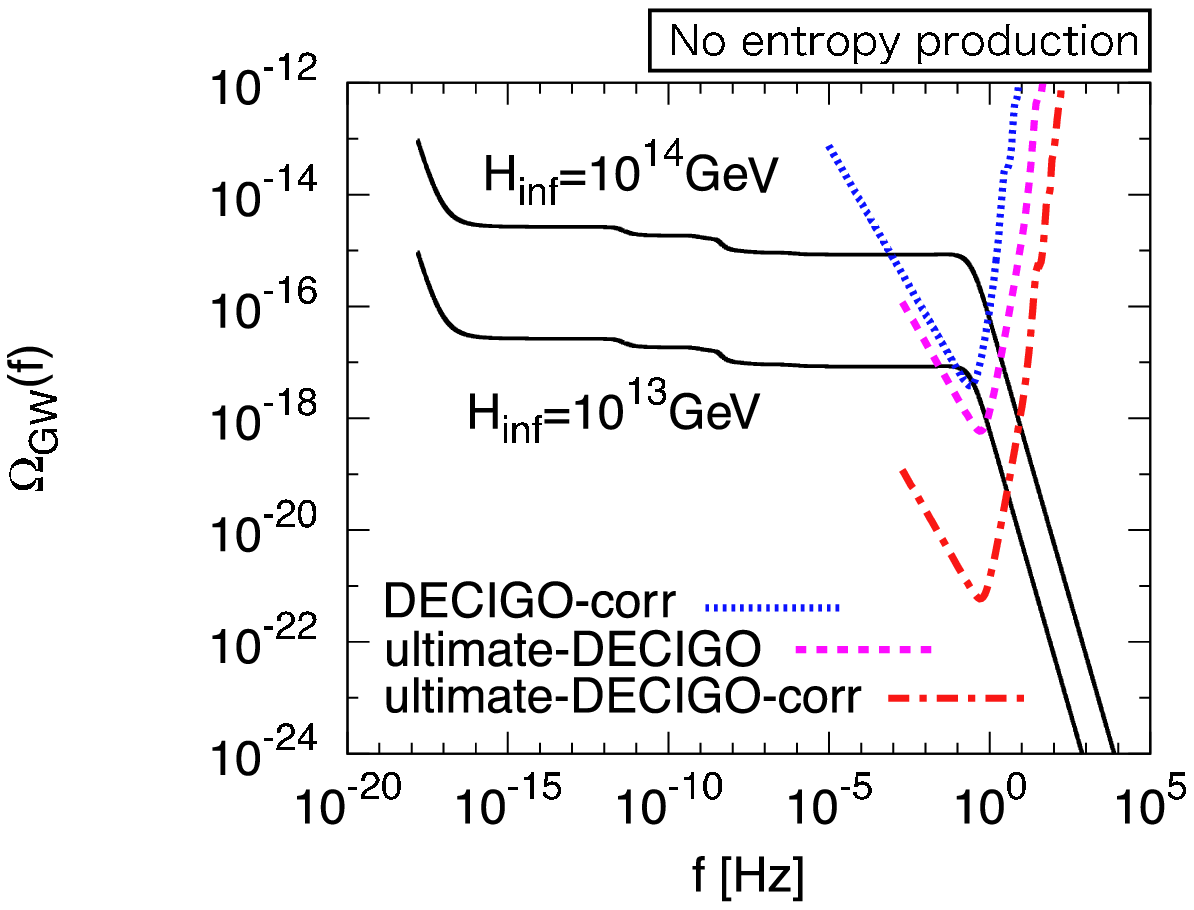}
  \includegraphics[width=0.7\linewidth]{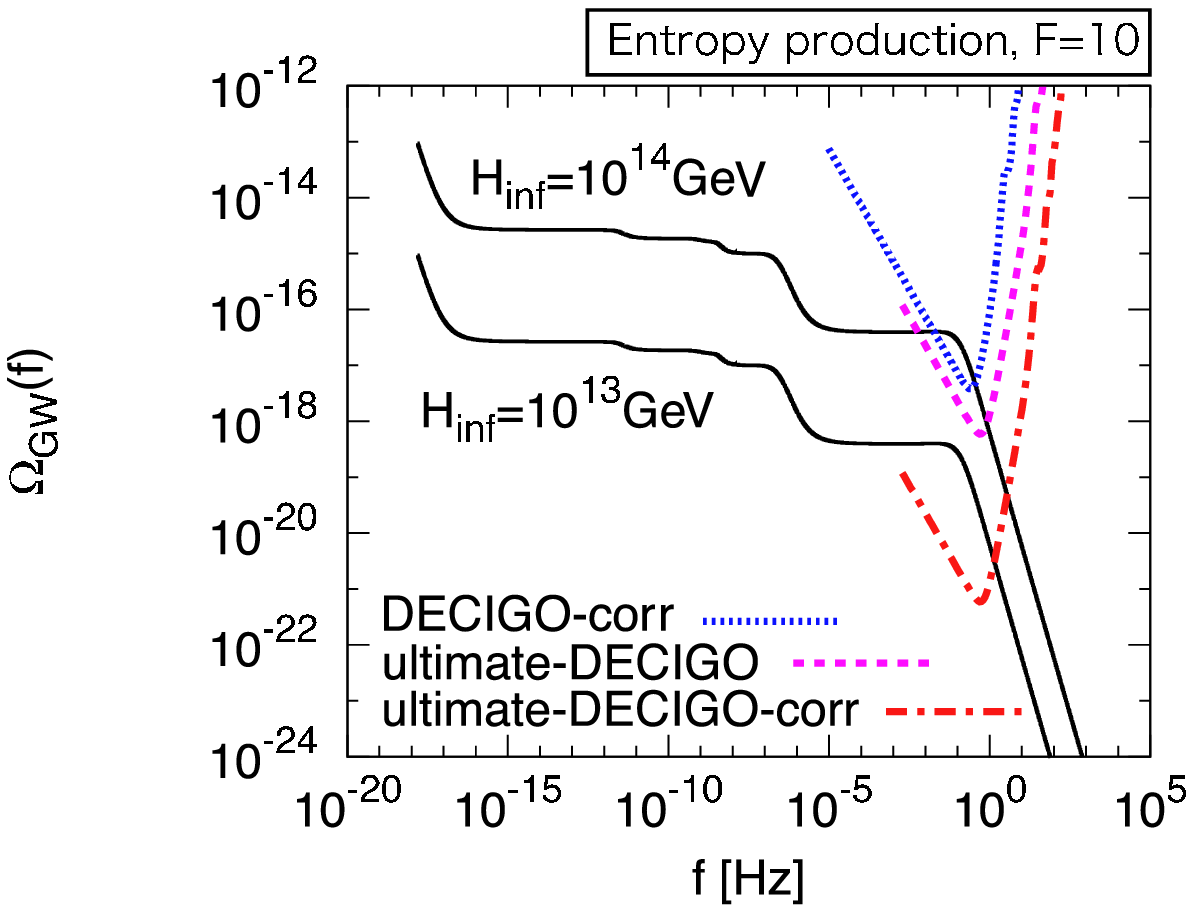}
  \caption{(Top) Spectra of the gravitational wave background for inflationary scale
  	$H_{\rm inf}=10^{14}$~GeV and $10^{13}$~GeV.
	Here we have taken $T_{\rm R}=10^7$~GeV.
	Also shown are sensitivities of planned space-based gravitational wave detectors,
	DECIGO with a correlation analysis (blue dashed line), 
	ultimate-DECIGO (purple dotted line), and correlation of analysis of
	ultimate-DECIGO (red dot-dashed line).
	(Bottom) Same as the top panel for the dilution factor $F=10$ for 
	$T_\sigma$=10~GeV and $T_{\rm R}=10^7$~GeV.
  }
   \label{fig:GWspec}
 \end{center}
\end{figure}

In Fig.~\ref{fig:GWspec} we show spectra of the 
gravitational wave background 
as a function of its present frequency.
In the top panel, the spectra for $H_{\rm inf}=10^{14}$~GeV 
and $10^{13}$~GeV
with the reheating temperature $T_{\rm R}=10^6$~GeV are shown.
Also plotted are sensitivities of
DECIGO with a correlation analysis (blue dashed line), 
ultimate-DECIGO (red dotted line), and correlation of analysis of
ultimate-DECIGO (purple dot-dashed line) \cite{Kudoh:2005as}.
In the bottom panel, the gravitational wave background 
spectra in the presence of entropy production
is shown, for $F=10$ and $T_\sigma$=10~GeV and $T_{\rm R}=10^7$~GeV.


\subsection{Scalar perturbation in the curvaton scenario}

The curvaton is a scalar field other than the inflaton, 
which remains light during inflation 
and has a quantum fluctuation $\delta \sigma \sim H_{\rm inf}/(2\pi)$.
Then, according to the $\delta N$-formalism \cite{Starobinsky:1986fxa}, 
the comoving
curvature perturbation $\zzeta$ is given by
\begin{equation}
	\zzeta = N_\sigma \delta \sigma 
+\frac{1}{2}N_{\sigma \sigma}(\delta \sigma)^2,
\end{equation}
up to the second order in $\delta \sigma$,
where $N$ is the local number of $e$-folds, given by 
the integral of the local expansion from
an initial spatially flat hypersurface to a final 
uniform density hypersurface,
and $N_\sigma$ is its derivative with respective to $\sigma$.
Note that we assume the curvaton behaves as free scalar 
field and hence
$\delta \sigma$ can be regarded as a random Gaussian variable.
Then we obtain \cite{Lyth:2005fi}
\begin{equation}
	\zzeta = \frac{2R}{3}\left ( \frac{\delta \sigma}{\sigma_i} 
\right )
	+\left( \frac{R}{3}-\frac{4R^2}{9}-\frac{2R^3}{9} \right)
	\left ( \frac{\delta \sigma}{\sigma_i} \right )^2,  \label{zeta}
\end{equation}
where $\sigma_i$ is the initial amplitude of the curvaton 
during inflation, and
\begin{equation}
	R=\left. \frac{3\rho_\sigma}{4\rho_r + 3\rho_\sigma} 
 \right|_{\sigma~{\rm decay}}
\end{equation}
roughly denotes the fraction of the curvaton energy density 
to the total energy density 
at the epoch of curvaton decay.
Here $\rho_\sigma$ and $\rho_r$ are the energy densities of 
the curvaton and that of radiation, respectively.
In order to reproduce the observed density perturbation of the Universe,
we must have
\begin{equation}
	\sqrt{ \Delta_\zzeta^2(k_*)} =\frac{R}{3}
\left(\frac{H_{\rm inf}}{\pi \sigma_i}\right)
	= 5\times 10^{-5}.  \label{normal}
\end{equation}

Notice that the
 inflaton also generates curvature perturbation whose magnitude is
$\Delta_{\zzeta\phi}^2 = H_{\rm inf}^2/(8\pi^2 \epsilon M_{\rm Pl}^2)$.
It is much smaller than curvaton's contribution (\ref{normal})
by assumption.
This leads to a constraint
\begin{equation}
	\epsilon > \frac{9}{8R^2}\left ( \frac{\sigma_i}{M_{\rm Pl}}
 \right )^2
	= 9\times 10^{-3} \left ( \frac{H_{\rm inf}}{10^{14}
~{\rm GeV}} \right )^2, \label{epsconst}
\end{equation}
where we have used the WMAP normalization (\ref{normal}).
We can calculate the tensor-to-scalar ratio $r$ in the curvaton scenario, as
\begin{equation}
	r \equiv \frac{\Delta_h^2(k_*)}{\Delta_\zzeta^2(k_*)}
	=\frac{18}{R^2}\left ( \frac{\sigma_i}{M_{\rm Pl}} \right )^2
 =0.14\left(\frac{H_{\rm inf}}{10^{14}\mathrm{GeV}}\right)^2.
\label{r}
\end{equation}
It can be checked that this is smaller than $16\epsilon$ using 
Eq.~(\ref{epsconst}), 
which is the prediction of the standard inflation scenario, 
once we assume that the curvature perturbation 
from the inflaton should be smaller than that from the curvaton.

The scalar spectral index is given by \cite{Lyth:2001nq}
\begin{equation}
	n_s = 1-2\epsilon +\frac{2m_\sigma^2}{3H_{\rm inf}^2},
\end{equation}
where $m_\sigma$ is the curvaton mass.
We can determine it through the relation
\begin{equation}
m_\sigma^2=\frac{3}{2}(n_s-n_t-1)H_{\rm inf}^2
=11(n_s-n_t-1)r\times(10^{14}\mathrm{GeV})^2,
\end{equation}
in principle.  But it is generically much smaller than
$H_{\rm inf}^2$, so that it would be more practical to use
the above equality as a consistency relation,
\begin{equation}
n_s-n_t\cong 1,  \label{consistency}
\end{equation}
 for the curvaton scenario.

Thus in the curvaton scenario, both the scalar and tensor 
spectral tilt are predicted to be
$-2\epsilon$, and hence a measurement of the tensor spectral index by
future space-based gravitational wave detectors~\cite{Seto:2005qy}
will give an evidence of the curvaton scenario.

\subsection{Non-Gaussianity in the curvaton scenario}

Statistics of the observed CMB anisotropy is currently consistent 
with Gaussian distribution.
The deviation from Gaussianity is parameterized by the non-linearity 
parameter $f_{\rm NL}$,
whose definition is given by
\begin{equation}
	\zzeta = \zzeta^{(\rm g)}+\frac{3}{5}f_{\rm NL}\zzeta^{(\rm g)2},
\end{equation}
where $\zzeta^{(\rm g)}$ denotes the Gaussian part of the curvature 
perturbation.
WMAP5 result gives a constraint on it as $-9<f_{\rm NL}<111$ at 95\% 
C.L. \cite{Komatsu:2008hk}.
From Eq.~(\ref{zeta}), we can estimate the non-linearity parameter 
as\footnote{
	Here we assume that there are no CDM/baryonic isocurvature 
perturbations,
	which might cause other types of non-Gaussianity 
	\cite{Kawasaki:2008sn,Langlois:2008vk,Hikage:2008sk}.
}
\begin{equation}
	f_{\rm NL}=\frac{5}{4R}\left (1- \frac{4}{3}R-\frac{2}{3}R^2 
\right ).  \label{fNL-R}
\end{equation}
Thus $f_{\rm NL}$ can be significantly large for small $R$,
and this may provide an observational hint of the curvaton scenario
if large $f_{\rm NL}$ is detected,
because standard inflation models predict a small non-linearity 
parameter, 
$f_{\rm NL}\sim \mathcal O (\epsilon, \eta)$.
Here we have assumed that the curvaton potential is quadratic, 
$V=(1/2)m_\sigma^2 \sigma^2$.
If it deviates from the quadratic one, the prediction of 
$f_{\rm NL}$ changes \cite{Enqvist:2005pg}
and in this case the coefficient of the trispectrum, $g_{\rm NL}$, 
may be useful to distinguish curvaton models.
We do not go into the detail on this point.

Notice that in the case where a large non-Gaussianity 
($f_{\rm NL} \gtrsim 10$) is obtained,
the curvaton decay does not increase entropy because 
it must be a subdominant component at the instance 
of its decay ({\it i.e.}, $R\ll 1$),
so that the gravitational wave background spectrum is not modified.
In this sense, the detections of $f_{\rm NL}$ by the CMB observation
and entropy production process by the gravitational wave 
background as a probe of the curvaton
scenario are complementary to each other.

To summarize, we have five possible observable quantities: scalar spectral index $n_s$,
tensor-to-scalar ratio $r$, tensor spectral index $n_t$, 
the dilution factor $F$ and non-linearity parameter $f_{\rm NL}$.
Among them, $r$ will be accurately determined by future B-mode polarization measurements.
Direct detection of gravitational waves will observe $n_t$, and comparing it with $n_s$
will confirm the curvaton scenario.
Then, either large enough $F$ or $f_{\rm NL}$ will be observed depending on whether
the curvaton once dominated the Universe or not.
In the former case, $F$ can be determined
by comparing $r$ and directly observed magnitude of the gravitational waves.
In the latter case, $f_{\rm NL}$ will be determined by CMB measurements such as Planck.
Then these observables may be used to pin down a curvaton model.

In the next section we investigate the possibility to 
detect either signature of the curvaton scenario and how they can
fix properties of the curvaton.

\section{Probing the Curvaton Scenario} \label{sec:probe}

We have seen that a curvaton scenario may leave distinct signatures 
on either a shape of the gravitational wave 
background or primordial non-Gaussianity.
Here we show parameter regions where a curvaton scenario has 
characteristic features on either of them.

Curvaton models are characterized by two parameters, 
the initial amplitude of the curvaton
$\sigma_i$ and its decay temperature $T_\sigma$.
The relevant quantity is the abundance of the curvaton 
coherent oscillation at the time of its decay,
which depends on these two parameters and the reheating 
temperature after inflation, 
$T_{\rm R}$, but does not depend on the curvaton mass.\footnote{
	If the curvaton begins to oscillate after the inflaton 
	decays, the mass dependence appears.
	In this case, however, it is sufficient to replace 
	$T_{\rm R}$ with $T_{\rm osc}$,
	the temperature at which the curvaton begins to oscillate, 
	as in Eq.~(\ref{F}).
}
Once these parameters are fixed, we can estimate the inflation 
scale $H_{\rm inf}$ in order to reproduce the observed magnitude of the density 
perturbation (see Eq.~(\ref{normal})).
The inflation scale $H_{\rm inf}$ 
also gives overall normalization of the 
gravitational wave background spectrum,
which can be directly measured independently using B-mode
polarization of CMB.  The spectral 
shape is determined by $T_{\rm R}$ and the curvaton 
abundance at its decay.
Thus we can uniquely predict the gravitational wave signal 
at an arbitrary frequency 
for each parameter set ($T_{\rm R}, T_{\sigma}, \sigma_i$).
Similarly, the level of non-Gaussianity depends only on the 
curvaton abundance
at its decay, as given in Eq.~(\ref{fNL-R}), and hence is 
also uniquely predicted.

Figures 2 depict the parameter region of the curvaton
which can be probed by observation of DECIGO/BBO.
The upper panel represents the region
accesible with a single ultimate-DECIGO 
and B-mode experiments such as EPIC \cite{Bock:2008ww}, 
CMBPol \cite{Baumann:2008aq}, and 
LiteBIRD \cite{LiteBIRD}
which are expected to reach down to $r \simeq 10^{-3}$.
The lower panel shows the ideal case of
correlation analysis of ultimate-DECIGO together with low-noise
delensed CMB map which would hopefully reach $r\sim 2\times 10^{-6}$
\cite{Marian:2007sr} (see also \cite{Knox:2002pe}).

Sensitivities of these direct detection experiments for the low frequency
are limited by stochastic noise from white dwarf binaries 
\cite{Farmer:2003pa},
and hence we have cut the sensitivities below 0.1~Hz.
We have set $\epsilon$ to satisfy the constraint (\ref{epsconst}), 
but the precise value of $\epsilon$
does not affect the results as long as it is sufficiently small.
The region above the purple wedge corresponds to $r>0.2$ which
is excluded by WMAP.

Also plotted there are contours of the non-linearity parameter
$f_{\rm NL}$ and the dilution factor $F$.  The region with
$f_{\rm NL}> 100$ is also disfavored by WMAP.  On the other hand,
Planck can measure it if it lies in the range $10< f_{\rm NL}$.
As argued before, it occupies a detached domain from the region
with $F>1$.  Therefore below we study how the curvaton parameters
are determined in each domain separately.

First let us consider the case $f_{\rm NL}\gg 1$ is confirmed by, say,
Planck experiment.  Then from (\ref{fNL-R}) we find 
$R \cong 5/(4f_{\rm NL})$ and therefore (\ref{r}) determines the
initial amplitude of the curvaton in terms of the observable 
quantities alone as
\begin{equation}
 \frac{\sigma_i}{M_{\rm Pl}}=\frac{5}{12f_{\rm NL}}
\left(\frac{r}{2}\right)^{1/2}.  \label{sigmai}
\end{equation}
In this case the curvaton decays before dominating the cosmic
energy density.  Hence we find
\begin{equation}
	\left. \frac{\rho_\sigma}{\rho_r}\right|_{\sigma~{\rm decay}}
	=\frac{1}{6}\left(\frac{\sigma_i}{M_{\rm Pl}}\right)^2
	\frac{a(T_\sigma)}{a(T_{\rm R})}
	=\frac{1}{6}\left(\frac{\sigma_i}{M_{\rm Pl}}\right)^2
	\frac{T_{\rm R}}{T_\sigma}\cong \frac{4}{3}R.
\end{equation}
These two equalities lead to
\begin{equation}
	\frac{T_\sigma}{T_{\rm R}}=\frac{5r}{576f_{\rm NL}}.
\end{equation}
Thus both of the curvaton parameters are fixed by the observable 
quantities in this case.
Finally we note that in this case
there is a simple relation between $T_\sigma/T_{\rm R}$ and
$\sigma_i/M_{\rm Pl}$ as
\begin{equation}
  	\frac{T_\sigma}{T_{\rm R}}=\frac{f_{\rm NL}}{10}
	\left(\frac{\sigma_i}{M_{\rm Pl}}\right)^2,
\end{equation}
which was used to draw the contours of $f_{\rm NL}$
in Figures 2.

\begin{figure}[]
 \begin{center}
 \includegraphics[width=0.7\linewidth]{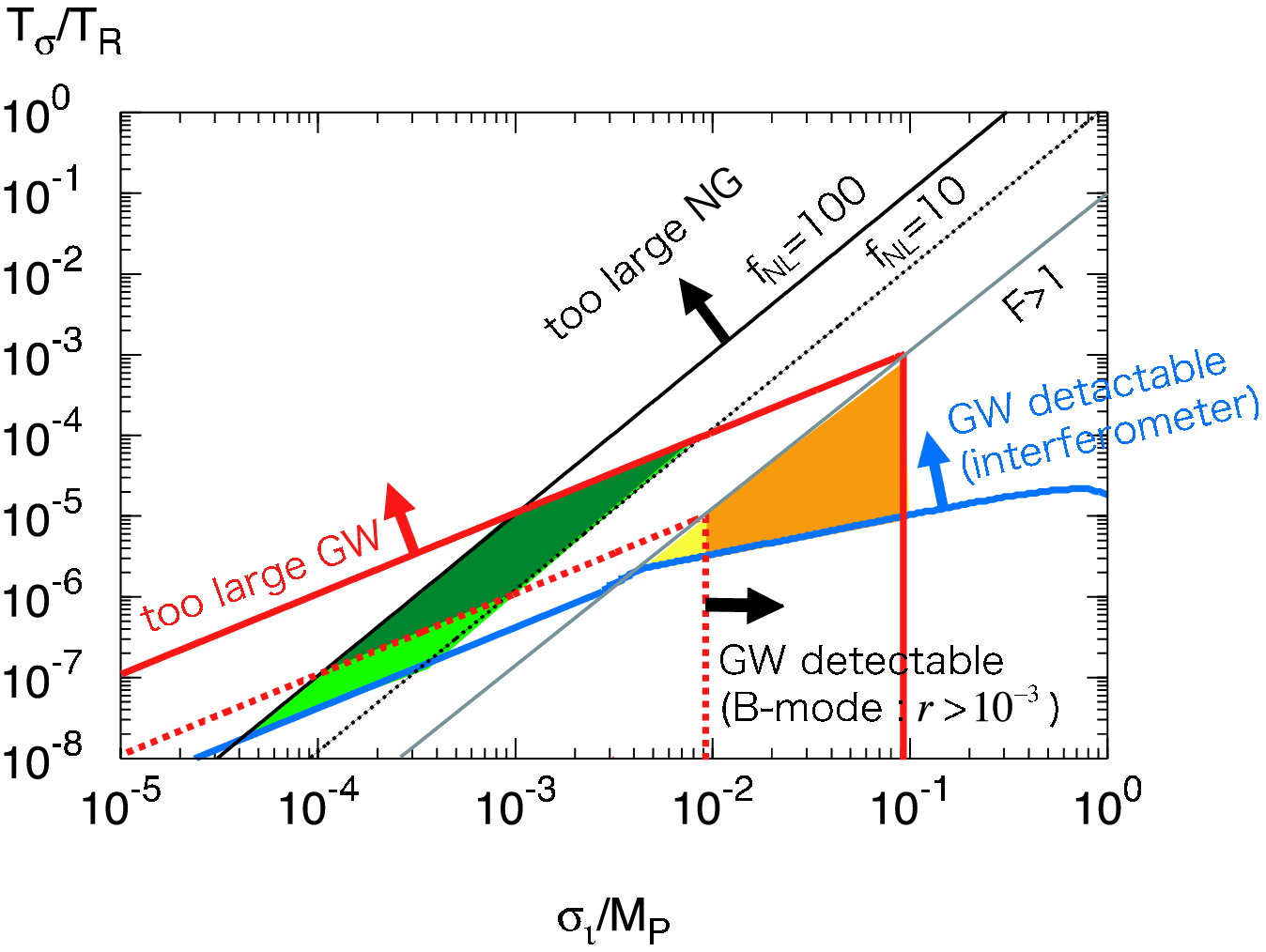}
  \includegraphics[width=0.7\linewidth]{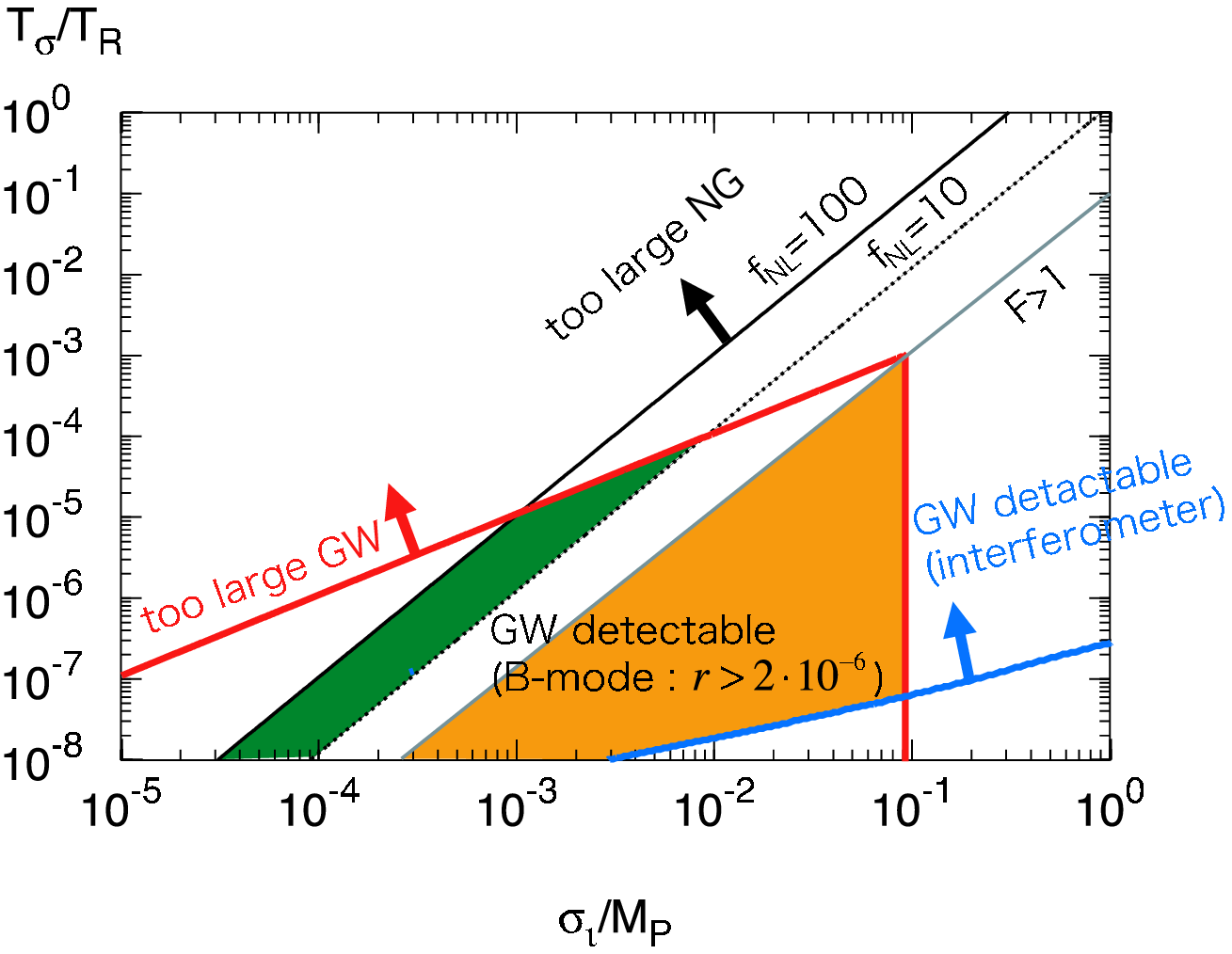}
  \caption{
Range of the curvaton parameters $\sigma_i/M_{\rm Pl}$-
  $T_\sigma/T_{\rm R}$
which can be probed by space-based
laser interferometers.  The upper panel represents the case with
 single ultimate-DECIGO and B-mode measurements down to $r=10^{-3}$,
while the lower panel shows an ideal case with 
 correlation analysis of
  ultimate-DECIGO and B-mode measurements accesible to
$r=2\times 10^{-6}$.
  Region above the red wedge 
is excluded since the tensor mode contribution to the 
  CMB anisotropy becomes too large.
 Also shown there are
  contours of the non-linearity parameter $f_{\rm NL}$.
Upper left region above the solid line is disfavored by WMAP.
In the green region all the curvaton parameters can be determined
in terms of $f_{\rm NL}$ and $r$, while in the orange region they can 
be determined by $F$ and $r$ provided other sources of entropy
production is absent.
In the yellow region, $F$ and $r$ are not determined independently, 
and only the combination $rF^{-4/3}$ is determined.
  }
   \label{fig:DEC}
 \end{center}
\end{figure}

Next we consider the case the curvaton dominates the energy density
of the universe when it decays releasing significant amount of
entropy with $R=1$ and $F > 1$.  Then from (\ref{r}) we find
\begin{equation}
	  \frac{\sigma_i}{M_{\rm Pl}}=\left(\frac{r}{18}\right)^{1/2}
\end{equation}
and from (\ref{F}) we obtain
\begin{equation}
  	\frac{T_\sigma}{T_{\rm R}}=\frac{r}{108F}, \label{tsigma}
\end{equation}
for $T_{\rm osc}>T_{\rm R}$.
We can determine $F$ if we can measure tensor perturbations both
by the B-mode polarization of CMB and DECIGO/BBO, provided there
is no other source of entropy production mechanism after reheating besides 
the curvaton.  Then all the relevant parameters are again fixed
by the observable quantities.
The corresponding parameter space is shown by the orange regions 
in Figs.~\ref{fig:DEC}.
In case there is other entropy production besides curvaton, 
the right-hand-side
of (\ref{tsigma}) will give a lower bound on $T_\sigma/T_{\rm R}$.
Finally, if B-mode polarization measurements fail to find 
the tensor mode and if it is detected only by DECIGO/BBO, which
is shown by the yellow regions in Figs.~\ref{fig:DEC},
the values of $F$ and $r$ cannot be determined independently.
Only the combination $rF^{-4/3}$ is determined in this case.

\if
Results are summarized in Fig.~\ref{fig:DEC}.
Contours of the non-linearity parameter $f_{\rm NL}$ are 
plotted on the $\sigma_i/M_{\rm Pl}$-
$T_\sigma/T_{\rm R}$ plane.
Regions where $f_{\rm NL}$ is much larger than 100 is 
inconsistent with WMAP observation
\cite{Komatsu:2008hk}.
On the other hand, if it lies in the range $10<f_{\rm NL}<100$, 
it can be confirmed by the Planck satellite.
Region above the purple line corresponds to $H_{\rm inf} > 
10^{14}$~GeV, which
is excluded since the tensor mode contribution to the CMB 
anisotropy becomes too large.
Future satellite experiments dedicated to detect a B-mode 
polarization of CMB
will find a primordial tensor mode for $H_{\rm inf}\gtrsim 10^{13}~$GeV
\cite{Bock:2006yf}.
Also we show a region where space-laser interferometer experiments 
can detect a stochastic gravitational wave background by the blue line.
Top and bottom panels correspond to single ultimate-DECIGO and
correlation analysis of ultimate-DECIGO, respectively.

In the region $f_{\rm NL} \ll 10$ and above the detectability 
lines of gravitational waves,
a significant amount of entropy production occurs,
which can be confirmed by these gravitational wave detectors.
It can be seen that such a confirmation is complementary to the 
detection of non-Gaussianity
from these figures, as already stated.
The region below all of these lines is also allowed, although no observable signature may arise.
\fi


\section{Conclusion} \label{sec:conc}

In this paper we have investigated a possible observable signatures from
curvaton scenarios, including non-Gaussianity, detection of tensor 
modes in CMB and direct detection of gravitational wave background.
First the measurement of the tensor power spectrum by CMB
is very important to identify the curvaton scenario through
the consistency relation (\ref{consistency}).
If the curvaton once dominates the Universe,
an entropy production process by its decay is imprinted in
the gravitational wave background spectrum and can be confirmed by future
space-based laser interferometer experiments.
For the opposite case where the curvaton is a subdominant component 
before it decays,
a large non-Gaussianity is predicted, which may be confirmed by 
Planck experiment.
In this sense, the gravitational wave signal and non-Gaussian signal 
are complementary to each other.
This allows us to probe the curvaton scenario for large parameter spaces,
giving information on the properties of the curvaton such as its 
decay rate and amplitude, etc.
It will in turn provide important information on high-energy physics beyond 
the reach of terrestrial experiments.

\section*{Acknowledgements}


K.N. would like to thank S.~Kuroyanagi for valuable discussion.
He would also 
like to thank the Japan Society for the Promotion of Science
for financial support.  This work was supported in part by
JSPS Grant-in-Aid for Scientific 
Research No.\ 19340054(JY) and
by Global COE Program ``the Physical Sciences Frontier", MEXT, Japan.


{}

\end{document}